\documentclass[9pt,a4paper]{article}
\usepackage{amsmath,amstext,amssymb,indentfirst,latexsym}
\usepackage{graphicx}

\newcommand{\mbi}[1]{\mbox{\boldmath $#1$}}

\def\cl#1{{\cal#1}}
\def\ket#1{|#1\rangle} \def\bra#1{\langle#1|}

\def\tr{{\rm Tr}}
\def\tran{\,^t}

\newtheorem{definition}{Definition}
\newtheorem{proposition}{Proposition}
\newtheorem{lemma}{Lemma}
\newtheorem{theorem}{Theorem}
\newtheorem{aproblem}{Problem}

\newtheorem{remark}{Remark}

\begin{document}

\title{Quantum estimation for non-differentiable models}
\author
{Yoshiyuki Tsuda${}^{1}$ and Keiji Matsumoto${}^{2,3}$}
\date
{
	${}^1$COE, Chuo University,
	1-13-27 Kasuga, Bunkyo-ku, Tokyo 112-8551, Japan.\\
	${}^2$National Institute of Informatics,
	2-1-2 Hitotsubashi, Chiyoda-ku, Tokyo 101-8430,
	Japan.\\
	${}^3$Imai Quantum Computation and Information Project,
	ERATO, JST, 5-28-3-201 Hongo, Bunkyo-ku, Tokyo 113-0033,
	Japan.
}
\maketitle

\begin{abstract}
State estimation is a classical problem in quantum information.
In optimization of estimation scheme, 
to find a lower bound to the error of the estimator is a very
important step.
So far, all the proposed tractable lower bounds use
derivative of density matrix.
However, sometimes, we are interested in
quantities with singularity, e.g. concurrence etc.
In the paper, lower bounds to a Mean Square Error (MSE) of an estimator are derived for a quantum estimation problem
without smoothness assumptions.
Our main idea is to replace the derivative by difference,
as is done in classical estimation theory.
We applied the inequalities to several examples,
and derived optimal estimator for some
of them.
\end{abstract}

\section{Introduction}
The quantum state estimation is a classical problem
in quantum information (\cite{helstrom, holevo}).
It is not only useful for various purposes,
e.g., evaluation of realized
quantum information processing systems, 
but also is a fundamental problem in its own right.

As is the case in classical estimation theory
\footnote{Throughout the paper, the estimation theory
of classical probability
distributions is called 'classical estimation'.},
often, it is assumed that the unknown source is a member of
a parameterized family of quantum states,
which is called {\it model}, 
to reduce the problem to the estimation of the unknown 
(in general, vector valued) parameter.
Sometimes, we are interested 
in some physical quantities of the state, e.g. entanglement of the state.
In such cases, the estimation of a function of
the unknown parameter is considered.

So far, most of the studies of quantum estimation theory
assumed that the model is a smooth surface in
the space of quantum states,  and that 
the function to be estimated are differentiable
up to necessary order.
Under these smoothness assumptions, 
tractable lower bounds to the MSE of 
the estimator had been studied, and some of them give
achievable bounds (\cite{helstrom, holevo, yuenlax}). 
Especially, in asymptotic regime,
we have lots of results on achievable bounds.

On the other hand, when 
there are singularities 
in the model and/or in the function to be
estimated,
a little had been
found out, so far as general theory is concerned.
It should be stressed that there are
some practically important examples to which 
the smoothness assumptions do not apply.
For example, concurrence, a measure of
entanglement, is singular at zero point.
Also, considering that 
the state estimation is
a fundamental problem of quantum information, 
it is better to be studied 
in most general setting.

The discrete models, or the models whose parameter
takes values in a discrete set, 
are only non-differentiable
models which has been studied intensively,
and there are many results in 
special cases, mainly by exploiting symmetry.
However, there are relatively a few  results 
which are valid for all discrete models.
To begin with,
it is not straightforward
even to work out a non-trivial lower bound to the error
of the optimal measurement.
It is true that  the formula in
\cite{YuenKennedyLax, Holevo73},
which is a special case of complementarity theorem
of SDP,
gives necessary and sufficient condition
for the optimal measurement.
However, this formula is far from tractable in many cases,
for being a system of non-linear
matrix equations and inequalities.

The purpose of the paper is to give the first 
general treatment of quantum state estimation
without the smoothness assumptions,
developing quantum versions of Hammersley-Chapman-Robbins inequality
(\cite{hammersley}, \cite{chapmanrobbins}),
Kshirsagar inequality (\cite{kshirsagar}) and
Koike inequality (\cite{koike}) in classical statistics.
While the lower bounds under the smoothness assumption
uses differential of density matrix, our new inequalities
use difference.

The paper is organized as follows. 
The Section \ref{sec:pre} is preliminaries. 
We give a brief review on classical and quantum
estimation theory, and set up the problem.
In Section \ref{sec:hcrk}, we state our new inequalities, whose proofs are given in
the Section \ref{sec:proof}.
In the Section \ref{sec:example}, our theory is applied to several
examples.
Non-asymptotic and asymptotic inequalities are derived, 
and based on those, optimal estimate are given for some of the examples.
In Section \ref{sec:discuss}, we state the conclusions
and the future problems.

\section{Preliminaries}
\label{sec:pre}
\subsection{Classical estimation theory}
An important problem of classical estimation is to optimize
estimator of (a function of) the parameter which is assigned
for the unknown probability distribution.
The error is often evaluated by MSE under unbiased condition.
Here, the estimator is said to be unbiased 
if its expectation equals the true value of (a function of)
the parameter.

In many cases, the optimal solution is too hard to obtain.
However, a series of tractable
lower bounds to MSE's of unbiased estimators are derived
using the first order and the higher order derivatives
of the probability distribution with respect to the parameter
of the model.
The inequality based on the first order 
derivative is called Cram\'er-Rao (CR) inequality (\cite{lehmann}),
while those which exploit also the higher order derivatives 
are called Bhattacharyya inequalities (\cite{bhattacharyya}).
By definition, Bhattacharyya inequality is always
not worse than CR inequality.

In general, none of these lower bounds is achievable, but
yet there are many cases where one of them can be achieved by
some estimators.
In asymptotic limit, 
an asymptotic CR inequality gives achievable lower bounds in general
(\cite{akahiratakeuchiasymptotics}). 
Needless to say, these inequalities rely on the smoothness assumptions.

These inequalities can be generalized to the cases with singularity 
by replacing (resp. higher order) derivatives by (resp. higher order) differences.
A difference version of CR inequality (resp. Bhattacharyya inequality)
is the Hammersley-Chapman-Robbins-Kshirsagar (HCRK) inequality
(\cite{hammersley}, \cite{chapmanrobbins}, \cite{kshirsagar})
(resp.
Koike inequality (\cite{koike})).
By definition, Koike inequality is always
not worse than HCRK inequality.
When the singularity is absent,
by taking difference infinitesimally small,
those difference versions reproduces the derivative versions.
We can also make 
an asymptotic version of HCRK inequality, but 
unlike the differentiable cases,
this bound is not always achievable (\cite{shibata}).
See \cite{akahiratakeuchinonregular} for more
detailed description of the theory which allows singularity.

\subsection{Quantum estimation theory with smoothness assumptions (I)}
\label{subsec:q-est}
Quantum versions for inequalities under the smoothness assumptions, i.e.,
Quantum Cram\'er-Rao (QCR) inequality
(\cite{helstrom}, \cite{holevo}, \cite{yuenlax})
and Quantum Bhattacharyya (QB) inequality
(\cite{brodyhughston}),
have already been studied a lot.

One important point is that,
due to the non-commutativity of quantum mechanics,
there are many quantum versions of those inequalities
(Quantum Cram\'er-Rao (QCR) inequalities).
A difficulty is that there is no 'best' QCR inequality.
That is, to obtain the tighter lower bound,
one must chose a proper version of QCR 
depending on the estimation problem of interest.

However, if either the parameter of the model is scalar valued,
or a scaler valued function of the parameter is to be estimated,
an SLD QCR (defined later) gives tighter lower bound than
any other QCR's.
Also, in this case, 
an SLD QCR gives the asymptotically achievable lower bound to the error of
estimators, as in the classical estimation theory.

Now, we state the mathematical detail of the theory.
Let $\cl H$ be a separable Hilbert space
describing the physical system of interest.
Denote 
the set of linear operators on
$\cl H$
and  the set of density operators, 
by  $\cl L(\cl H)$ and 
by $\cl S(\cl H)(\subset\cl L(\cl H))$,
respectively.
Let $\Theta$ be a parameter space which is a subset of
the set $\mathbb R$ of real numbers,
and assume that the density matrices for the system of interest
is a member of  a parameterized family of states, 
$\{\rho_\theta ; \theta\in \Theta\}$, which is called a {\it model}.
Throughout the paper, except as otherwise noted, we assume
\begin{align}
 \rho_{\theta}>0,
\label{ass:inverse}
\end{align}
or the density matrix $\rho_\theta$ has its reverse.

The model is said to be differentiable
(resp. non-differentiable),
if the map $\theta\mapsto\rho_\theta$ is differentiable 
(resp. not differentiable).
We would like to estimate $g(\theta)$ by measuring the system
where $g(\theta)$ is a function of $\theta$.
An estimator of $g(\theta)$ is a measurement 
whose output is an estimate, {\it i.e.},
a POVM $M$ which 
takes values on $g(\mathbb{R})$.

\subsubsection*{Non-asymptotic theory with the smoothness assumptions}
In non-asymptotic setting,
we often impose unbiasedness condition on estimators, which
is defined as follows.

\begin{definition}
\rm
The bias $b(M,\theta)$ of an estimator $M$ at $\theta$ is
defined by
\[
	b(M,\theta)=\int_{\gamma\in g(\Theta)}
	(\gamma-g(\theta))\tr(\rho_\theta M(d\gamma)).
\]

If $b(M,\theta)\equiv0$,
$M$ is said to be {\it unbiased at $\theta$}.
If $M$ is unbiased at any $\theta$,
$M$ is said to be {\it unbiased}.
\end{definition}
The classical CR inequality uses 
a {\it logarithmic derivative} $l_{\theta}(x)$ which is defined as,
\[
 l_{\theta}(x):=\frac{d}{d\theta}\log p(x,\theta),
\]
or as the solution  to the equation
\[
 p(x,\theta)l_{\theta}(x)=
\frac{dp(x,\theta)}{d\theta}.
\]

There are many quantum analogues of a logarithmic derivative,
due to the non-commutativity of quantum mechanics.
Among those, a Symmetric Logarithmic Derivative (SLD) and 
a Right Logarithmic Derivative (RLD)
are most frequently used.
\begin{definition}
\rm
An SLD $L^S_\theta\in\cl L(\cl H)$ is
defined as a solution to the equation
\[
	\frac{d}{d\theta}\rho_\theta
	=\frac{\rho_\theta L^S_\theta+L^S_\theta\rho_\theta}{2},\
	L^S_\theta=(L^S_\theta)^\dagger
	.
\]
An RLD $L^R_\theta\in\cl L(\cl H)$
defined as a solution to the equation
\begin{equation}
	\frac{d}{d\theta}\rho_\theta
	=\rho_\theta L^R_\theta
	.
	\label{eq:def:rld}
\end{equation}
\end{definition}

An SLD and an RLD are uniquely defined since $\rho_\theta$ is
invertible by the assumption (\ref{ass:inverse}).
The performance of an estimator $M$ of $g(\theta)$ is evaluated by the MSE
\[
	V_\theta(M)=
	\int_{\gamma\in g(\Theta)}
	(\gamma-g(\theta))^2
	\tr(\rho_\theta M(d\gamma))
	.
\]
If the bias $b(M,\theta)$ is exactly zero, MSE is equal to what is called
variance.
An SLD and an RLD Quantum Cram\'er-Rao (QCR) inequality are stated as follows.
\begin{proposition}({\rm\cite{helstrom}}) \
If an estimator $M$ is unbiased,
it holds that
\begin{equation}
	\label{eq:qcr}
	V_\theta(M)
	\ge
	(g'(\theta))^2/J^S_\theta
	\ge
	(g'(\theta))^2/J^R_\theta
\end{equation}
where
\[
	J^S_\theta=\tr(\rho_\theta (L^S_\theta)^2),\
	J^R_\theta=\tr(\rho_\theta L^R_\theta (L^R_\theta)^\dagger).
\]
The equality in the first inequality in (\ref{eq:qcr}) 
holds if and only if
$M$ gives the spectral decomposition of a hermitian matrix $T$ such that
$T-g(\theta) I\in{\rm span}\{L^S_{\theta}\}$,
where $I$ is the identity on $\cl H$.
\end{proposition}
The proof of the proposition is reviewed in Section \ref{sec:proof}.

See \cite{lehmann} for classical version of the theory,
\cite{helstrom, holevo, yuenlax} for the quantum theory.

\subsubsection*{Asymptotic theory with the smoothness assumptions}
Assuming that $n$ identical independent samples 
$\rho_\theta^{\otimes n}$ of unknown states 
$\rho_\theta$ is given,
let us analyze the behavior of estimators as $n$ tends to infinity.
An estimator is a POVM $M$ living in $\cl H^{\otimes n}$.
If the smoothness assumptions hold,      
the mean square error converges to zero with the order of $O(1/n)$.
Therefore, we focus on the coefficient of the $1/n$-term.

In asymptotic analysis,  the unbiasedness condition is replaced by
asymptotic unbiasedness defied as follows.
\begin{definition}
\rm
For the sequence $M_n$ of estimators of $g(\theta)$,
if the bias $b(M_n,\theta)$ satisfies
\[
	b(M_n,\theta)=o(1/\sqrt n),\
	\left.\frac{d}{d x}b(M_n,x)\right|_{x=\theta}=o(1/\sqrt n),
\]
$M_n$ is said to be {\it asymptotically unbiased at
$\theta$}.
If $M_n$ is unbiased at any $\theta\in\Theta$,
$M$ is said to be {\it asymptotically unbiased}.
\end{definition}
The Asymptotic Quantum Cram\'er-Rao (AQCR) inequality
is stated as follows.
\begin{proposition}\label{propo:AQCR}
(\cite{Nagaoka:1989:2}, 
\cite{hayashimatsumotofreedom} 
\cite{gillmassar})
Assume that $M_n$ is asymptotically unbiased.
Then, we have
\begin{equation}
	\label{eq:AQCR}
	\lim_{n\to\infty} n V_\theta(M_n)
	\ge
	(g'(\theta))^2/J^S_\theta
	\ge
	(g'(\theta))^2/J^R_\theta
	.
\end{equation}
Especially, the first inequality is achieved by some 
asymptotically unbiased estimators.
\end{proposition}

\subsection{Quantum estimation theory with smoothness assumptions (II)}
In this subsection, we review quantum estimation theory
for vector valued parameters with smoothness assumptions.
For simplicity, here 
we concentrate on the estimation of the parameter itself,
and do not treat the estimation of $g(\theta)$.

Let us denote by $m$ the dimension of the parameter 
$\theta=(\theta^1,\cdots\theta^m)$
($m=1$ if $\theta$ is scalar), and define 
SLD $L^S_{i,\theta}$ of $\theta^i$ 
and 
RLD $L^R_{i,\theta}$
as the solution to the equations
\begin{align*}
\frac{\partial}{\partial\theta^i}\rho_{\theta}
=\frac{1}{2}(L^S_{i,\theta}\rho_\theta + \rho_\theta L^S_{i,\theta})
=\rho_\theta L^R_{i,\theta}.
\end{align*}
Define also matrices $J^S_\theta$ and $J^R_\theta$ by,
\begin{align*}
J^S_{\theta,i,j}
=\frac{1}{2}
\tr\rho_\theta
 (L^S_{i,\theta}L^S_{j,\theta}+L^S_{j,\theta}L^S_{i,\theta})
  ,
\quad\quad\quad
 J^R_{\theta,i,j}
=\tr\rho_\theta
   L^R_{j,\theta}(L^R_{i,\theta})^{\dagger}
  ,
\end{align*}
respectively.
\begin{proposition} (\cite{holevo}) 
Assume that the estimator $M$ is unbiased
(for each $\theta^i$). Then,
for any real valued positive matrix $G$, we have,
\begin{align*}
  {\rm Sp} G V_\theta(M)
&\geq 
{\rm Sp} G(J^S_{\theta})^{-1},\\ 
 {\rm Sp} G V_\theta(M)
&\geq 
{\rm Sp} G(J^R_{\theta})^{-1}
+{\rm Spabs}\Im G(J^R_{\theta})^{-1}.
\end{align*}
\label{prop:qcr-multi}
\end{proposition}
The proof is reviewed in Section~\ref{sec:proof}.
Here, ${\rm Sp}$ is the trace over the vector space $\mathbb{C}^m$,
and ${\rm Spabs}A$ is the sum of absolute values of eigenvalues of $A$.
Typically, $G$ is chosen to be ${\rm diag}(g_1,...,g_m)$, and
in that case ${\rm Sp} G V_\theta(M)$ is a weighed sum of
MSE's of the estimators of components of $\theta$.

Different from the scalar valued parameter case,
we cannot say which one is better in general.
However, if $[L^S_{i,\theta}, L^S_{j,\theta}]=0$,
SLD QCR gives better lower bound, and is asymptotically
achievable by the estimator based on the simultaneous spectral
decomposition of SLD's. 
If LSD's are not commutative, RLD QCR will be often useful.

\section{New inequalities for non-differentiable models}
\label{sec:hcrk}
\subsection{Non-asymptotic theory without the smoothness assumptions}
If $\rho_\theta$ 
and/or the function $g(\theta)$ are
 not differentiable with respect to $\theta$,
then the inequality (\ref{eq:qcr}) does not make any sense.
In classical statistics, Hammersley \cite{hammersley} and
Chapman and Robbins \cite{chapmanrobbins} have
derived an inequality using the one-sided difference
in place of the derivative,
and
Kshirsagar \cite{kshirsagar} improved it by
the two-sided difference.
We generalize their idea to the quantum theory.
The proofs of the theorems are given in Section \ref{sec:proof}.

Let $\cl F$ be the linear space of
$\Bbb R$-valued functions of
$\theta\in\Theta$.
For $f(\theta)\in\cl F$
and for $\delta\in\Bbb R$,
let  $\Delta_\delta$ be a linear operator defined by
\[
	\Delta_{\delta}^t f(\theta)
	=
	\frac
	{f(\theta + t \delta)
	-
	f(\theta-(1-t)\delta)}
	{\delta}.
\]
We next define operators,
which are analogues of SLD and RLD.
\begin{definition}
\rm
For $\delta\in\Bbb R$,
define $L^{S,t}_{\theta,\delta}
(=(L^{S,t}_{\theta,\delta})^\dagger)$
and $L^{R,t}_{\theta,\delta}$ by
\begin{equation}
	\label{eq:lslrdelta}
	\Delta_{\delta}^t \rho_\theta
	=
	\frac
	{
	\rho_{\theta}L^{S,t}_{\theta,\delta}
	+L^{S,t}_{\theta,\delta}\rho_{\theta}
	}
	{2},\
	\Delta_{\delta}^t \rho_\theta
	=
	\rho_{\theta}L^{R,t}_{\theta,\delta}.
\end{equation}
($L^{S,t}_{\theta,\delta}$ and $L^R_{\theta,\delta}$
are uniquely defined since $\rho_\theta$ is
invertible by the assumption (\ref{ass:inverse}).)
\end{definition}
\begin{remark}\label{rem:sabun2bibun}
\rm
	If $\rho_\theta$ is differentiable with $\theta$,
	then
	$L^{S,t}_{\theta,\delta}\to
	L^S_{\theta}$
	and
	$L^{R,t}_{\theta,\delta}\to
	L^R_{\theta}$
	as $\delta\to0$
	since
	$\Delta_\delta^t \rho_\theta\to
	(d/d\theta)\rho_\theta$
	as $\delta\to0$.

If the model is differentiable  both from the left
and from the right,
we have, 
\begin{align}
\label{eq:16}
\lim_{\delta\to 0+}L^{S,t}_{\theta,\delta}
=t L^{S,+}_{\theta,t}+ (1-t)L^{S,-}_{\theta},\quad
\lim_{\delta\to 0+}L^{R,t}_{\theta,\delta}
=t L^{R,+}_{\theta,t}+ (1-t)L^{R,-}_{\theta},
\end{align}
where $L^{S,\pm}_\theta$ and $L^{R,\pm}_\theta$ are defined by,
\begin{align*}
 \lim_{\delta\to 0\pm}\Delta_{\delta}^1\rho_\theta
=\frac{1}{2}(\rho_\theta L^{S,\pm}_{\theta}+
 L^{S,\pm}_{\theta}\rho_\theta)
=\rho_\theta L^{R,\pm}_{\theta}.
\end{align*}
(Observe that 
\[
	\lim_{\delta\to+0}\Delta^t_{\delta}\rho_{\theta}=
	\lim_{\delta\to+0}
	t\Delta^1_{\delta}\rho_{\theta}
	+
	\lim_{\delta\to+0}(1-t)\Delta^0_{\delta}\rho_{\theta}.
\]
)
If the singular model of our concern is
embedded in the larger smooth model,
then $ L^{S,\pm}_{\theta}$ and $ L^{R,\pm}_{\theta}$
are calculated from SLD and RLD of the larger model.
\end{remark}
Our first result,
Quantum Hammersley-Chapman-Robbins-Kshirsagar (QHCRK) inequality,
is given in the following theorem.
\begin{theorem}\label{th:QHCR}
If the estimator $M$ of $g(\theta)$ is unbiased, then
\begin{eqnarray}
	V_\theta(M)
	&\ge&
	(\Delta_{\delta}^t g(\theta))^2
	/J^{S,t}_{\theta,\delta}
	\label{eq:qhcrequality}
	\\
	&\ge&
	(\Delta_{\delta}^t g(\theta))^2
	/J^{R,t}_{\theta,\delta}
	\label{eq:qhcrequality-r}
\end{eqnarray}
where
\[
	J^{S,t}_{\theta,\delta}=
	\tr(\rho_\theta (L^{S,t}_{\theta,\delta})^2),\
	J^{R,t}_{\theta,\delta}=
	\tr(\rho_\theta
	L^R_{\theta,\delta} (L^{R,t}_{\theta,\delta})^\dagger).
\]
The equality in (\ref{eq:qhcrequality}) holds if and only if
$M$ gives the spectral decomposition of a hermitian matrix $T$ such that
$T-\theta I\in{\rm span}\{L^{S,t}_{\theta,\delta}\}$.
\end{theorem}

We next present 
Quantum Koike (QK) inequality which improves QHCRK.
For a function $f(\theta)$ of $\theta\in\Theta$,
for a real number $\delta$ and
for an integer $k\ge1$,
we define 
$k$-th difference operator $\Delta_{\delta,k}$ by
\[
	\Delta_{\delta,k}f(\theta)
	=
	(-1)^k\frac1{\delta^k}
	\sum_{i=0}^{k}(-1)^i{k\choose i}f(\theta+\delta i)
	.
\]
Note that, if $f(\theta)$ is $k$-times differentiable, then
$\Delta_{\delta,k}f(\theta)\to
(d^k/d\theta^k)f(\theta)$ as $\delta\to0$.
Let us define
$L^S_{\theta,\delta,k}$ and $L^R_{\theta,\delta,k}$ as follows.
\begin{definition}
\rm
Define $L^S_{\theta,\delta,k}$ and
$L^R_{\theta,\delta,k}$ by
\begin{eqnarray*}
	\Delta_{\delta,k}\rho_{\theta}
	&=&
	\frac
	{\rho_\theta L^S_{\theta,\delta,k}
	+L^S_{\theta,\delta,k}\rho_\theta}{2}
	\quad
	(L^S_{\theta,\delta,k}=(L^S_{\theta,\delta,k})^\dagger),
	\\
	\Delta_{\delta,k}\rho_{\theta}
	&=&
	\rho_\theta L^R_{\theta,\delta,k}.
\end{eqnarray*}
\end{definition}
Let us define $r\times r$ matrices 
$K^S_{\theta,\delta}=(K^S_{\theta,\delta,i,j})$ 
and
$K^R_{\theta,\delta}=(K^R_{\theta,\delta,i,j})$ 

($i,j=1,...,r$) by
\[
K^S_{\theta,\delta,i,j}= 	
	\tr(\rho_\theta
	L^S_{\theta,\delta,i} L^S_{\theta,\delta,j}),\
K^R_{\theta,\delta,i,j}= 	
	\tr(\rho_\theta
	L^R_{\theta,\delta,i} (L^R_{\theta,\delta,j})^\dagger),
\]
and let
$\bold v=\tran(\Delta_{\theta,\delta,1}g(\theta),
	...,\Delta_{\theta,\delta,r}g(\theta))$ be a column vector.
We then have the following
Quantum Koike (QK) inequality which generalizes and
improves QHCRK inequality of
Theorem \ref{th:QHCR}.
\begin{theorem}\label{th:QK}
If an estimator $M$ for $g(\theta)$ is unbiased, then
\begin{eqnarray}
	\label{eq:th2:s}
	V_\theta(M)
	&\ge&
	\tran \bold v (K^S_{\theta,\delta})^{-1} \bold v,
	\\
	\label{eq:th2:r}
	V_\theta(M)
	&\ge&
	\tran \bold v (K^R_{\theta,\delta})^{-1} \bold v,
\end{eqnarray}
where $(K^S_{\theta,\delta})^{-1}$ and $(K^R_{\theta,\delta})^{-1}$ 
are generalized inverses.
The equality holds if and only if
$M$ can be written as an observable $T$ such that
$T-\theta I\in{\rm span}\{
L^S_{\theta,\delta,1},...,L^S_{\theta,\delta,k}\}$.
\end{theorem}

\begin{remark}\label{rem:koike2bhatt}
\rm
We can derive the Bhattacharyya inequality by
Theorem \ref{th:QK}.
\end{remark}

\subsection{Asymptotic theory without the smoothness assumptions}
If the smoothness assumptions are not valid,
MSE $V_\theta(M_n,\theta)$
of appropriate estimators $M_n$  does not necessarily
scale in $O(1/n)$.
(\cite{shibata} discusses this point in detail in the classical
estimation theory.)
The convergence rate was expressed by 
an increasing sequence $c_n$
that goes to infinity as $n\to\infty$.

In this paper, we study the following two cases.
\smallskip\\
{\bf Case 1} \
The parameter of the model takes discrete values,
and
each point in the model is isolated, 
and $g(\theta)=\theta$.
\\
{\bf Case 2} \
The model and the function $g(\theta)$ are continuous,
\smallskip\\
In Case 1,
if the model has only two density matrices,
the estimation of $\theta$ seems 
quite close to test of simple hypothesis about $\theta$.
We discuss this point later.
In both Cases 1 and 2,
we use an RLD-type inequalities, 
since SLD-type inequalities are mathematically intractable.

\subsubsection*{Discrete case}

If the set $\Theta$ of parameters is discrete
and each element of $\Theta$ is isolated
({\it i.e.}, 
there is a positive real number $c$ such that,
for any $\theta and \theta'\in\Theta$, it holds that
$\|\rho_\theta-\rho_{\theta'}\|_{tr}>c$),
then the convergence rates of the bias and the variance
of appropriate estimators are exponential.

Assume that
$\theta,\theta+\delta\in\Theta$, and that 
there is no element of $\Theta\subset\mathbb R$
between $\theta$ and $\theta+\delta$.
We also assume that $g(\theta)\ne g(\theta+\delta)$.
Then,
we have the following asymptotic QHCRK (AQHCRK) inequality
for the discrete case.
\begin{theorem}\label{th:AQHCRD}
Assume that there is $n_0$ such that, if $n\ge n_0$ then
$b(M_n,\theta)$ and $b(M_n,\theta+\delta)$ satisfy
\begin{equation}
	\label{eq:weakunbias}
	\frac{|b(M_n,\theta)|}{|\delta|}
	\le
	\frac{|\Delta_{\delta}g(\theta)|}{3}
	\mbox{, and }
	\frac{|b(M_n,\theta+\delta)|}{|\delta|}
	\le
	\frac{|\Delta_{\delta}g(\theta)|}{3}
	.
\end{equation}
Then
\begin{equation}
	\label{eq:c-ineq}
	\liminf_{n\to\infty}\frac1n\log(V_\theta(M_n))\ge
	-\log (1+\delta^2 J^{R,1}_{\theta,\delta}).
\end{equation}
\end{theorem}
{\bf Proof} \
Observe 
\begin{align}
 \tr\rho_{\theta}^{-1}\rho_{\theta+\delta}^{2}=
1+\delta^2 J^{R,1}_{\theta,\delta}.
\label{eq:rld}
\end{align}
Replacing $\rho_{\theta}$ and $\rho_{\theta+\delta}$ in (\ref{eq:rld}) by  
$\rho_{\theta}^{\otimes n}$ and $\rho_{\theta+\delta}^{\otimes n}$, 
we obtain,
\begin{align}
	\tr
	(
	\rho_\theta^{\otimes n}
	L^{R,(n)}_{\theta,\delta}
	(L^{R,(n)}_{\theta,\delta})^\dagger
	)
&= \frac{1}{\delta^2} 
   \big\{
         \tr(\rho_{\theta}^{\otimes n})^{-1}
         (\rho_{\theta+\delta}^{\otimes n})^{2}
         -1 
   \big\}
=  \frac{1}{\delta^2} 
   \big(
        (\tr\rho_{\theta}^{-1}\rho_{\theta+\delta}^{2})^{n}
         -1 
   \big) \nonumber\\
&= \frac{1}{\delta^2} 
   \big(
        (1+\delta^2 J^{R,1}_{\theta,\delta})^{n}
         -1 
   \big),
\label{eq:3:tohi}
\end{align}
where the last equation is due to (\ref{eq:rld}).

The assumption (\ref{eq:weakunbias})
and the formula (\ref{eq:proof:aqhcr})
imply that, for $n\ge n_0$,
\begin{eqnarray*}
	\frac 1 n \log V_\theta(M_n)
	&\ge&
	\frac 1 n
	2\log
	|\Delta_{\delta} b(M_n,\theta)+\Delta_{\delta}g(\theta)|
	-
	\frac 1 n
	\log
	\tr(\rho_{\theta,\delta}^{\otimes n}
	L^{R,(n)}_{\theta,\delta}(L^{R,(n)}_\delta)^\dagger)
	\\
	&=&
	\frac 1 n
	2\log
	|
	\frac{b(M_n,\theta+\delta)-b(M_n,\theta)}{\delta}
	+
	\Delta_{\delta}g(\theta)|
	-
	\frac 1 n
          \frac{(1+\delta^2 J^{R,1}_{\theta,\delta})^n-1} {\delta^2}
	\\
	&\ge&
	\frac 1 n
	2\log
	\frac{|\Delta_{\delta}g(\theta)|}{3}
	-
	\frac 1 n
	\log
          \frac{(1+\delta^2 J^{R,1}_{\theta,\delta})^n-1} {\delta^2}
	\\
	&\to&
	-\log (1+\delta^2 J^{R,1}_{\theta,\delta})
	\
	(n\to\infty)
	.
\end{eqnarray*}
Hence (\ref{eq:c-ineq}) holds.
\hfill$\Box$

\bigskip

We would like to remark that
$\log (1+\delta^2 J^{R,1}_{\theta,\delta})$
is always not smaller than
the relative entropy
$D(\rho_{\theta+\delta}\parallel\rho_\theta)$
defined as
\[
	D(\rho_{\theta+\delta} \| \rho_\theta)
	=
	\tr(\rho_{\theta+\delta}(\log\rho_{\theta+\delta}-
		\log\rho_\theta))
	.
\]
The reason is the following.
Consider a model $\Theta=\{\theta,\theta+\delta\}$.
Due to the theory of quantum hypothesis testing
(\cite{hiaipetz} says that \cite{ogawanagaoka}),
for any $\epsilon>0$,
there is a sequence $N_n$
of POVM's taking values in $\mathbb R$
such that
\begin{equation}
	\label{eq:hiaipetz}
	\tr(\rho_{\theta+\delta}^{\otimes n}
	N_n(\theta))
	\le
	\frac{|\Delta_\delta g(\theta)|}{3}
	,\
	\lim_{n\to\infty}\frac1n
	\log
	\tr(\rho_\theta^{\otimes n}N_n(\theta+\delta))
	=
	-D(\rho_{\theta+\delta}\parallel\rho_\theta).
\end{equation}
We can regard $N_n$ as an estimator of $\theta$
satisfying (\ref{eq:weakunbias}).
Indeed, there is $n_0$ such that, if $n\ge n_0$ then
\[
	\frac{|b(N_n,\theta)|}{|\delta|}
	=
	\frac
	{|\delta|\tr(\rho_\theta^{\otimes n}N_n(\theta+\delta))}
	{|\delta|}
	\le
	\frac{|\Delta_\delta g(\theta)|}{3}
	,\
	\frac{|b(N_n,\theta+\delta)|}{|\delta|}=
	\frac{|\delta|
	\tr(\rho_{\theta+\delta}^{\otimes n}
	N_n(\theta))}{|\delta|}
	\le
	\frac{|\Delta_\delta g(\theta)|}{3}
	.
\]
Moreover,
\begin{eqnarray*}
	\frac1n\log
	V_\theta(N_n)
	&=&
	\frac1n\log
	\big(
	(g(\theta+\delta)-g(\theta))^2
	\tr(\rho_{\theta}^{\otimes n}
	N_n(\theta+\delta))
	\big)
	\\
	&\to&
	-
	D(\rho_{\theta+\delta}\parallel\rho_\theta)
	\
	(n\to\infty).
\end{eqnarray*}
Due to (\ref{eq:c-ineq}),
we have
$D(\rho_{\theta+\delta} \| \rho_\theta)
 \le
\log (1+\delta^2 J^{R,1}_{\theta,\delta})
$.
\subsubsection*{Continuous case}
Let
$L^{R,n,t}_{\theta,\delta}$ be
the solution to the equation
\[
	\Delta_\delta^t \rho_\theta^{\otimes n}
	=
	\rho_\theta^{\otimes n}
	L^{R,n,t}_{\theta,\delta}
	.
\]
For $h\in\mathbb R$,
let $\delta=h/c_n$. Define that 
\begin{eqnarray*}
	J^{R,t}_\theta
	&=&
	\limsup_{h\to+0}
	\limsup_{n\to\infty}
	\frac{1}{c_n^2}
	\tr(\rho_{\theta}^{\otimes n}
	L^{R,n,t}_{\theta,h/c_n}
	(L^{R,n,t}_{\theta,h/c_n})^\dagger
	),
	\\
	g'_t(\theta)
	&=&
	\lim_{\delta\to+0}
	\Delta_{\delta}^t g(\theta)
	.
\end{eqnarray*}

If $\rho_\theta$ is differentiable from the right (resp. left),
$J^{R,1}_\theta$ (resp. $J^{R,0}_\theta$) is equal to $J^R_\theta$
where the two-sided derivative in (\ref{eq:def:rld})
is replaced with the right (resp. left) derivative.
This is shown as follows.
Replacing $\delta$ with $h/\sqrt n$ and $c_n$ with $\sqrt n$
in (\ref{eq:3:tohi}),
we have
\begin{eqnarray}
	\nonumber
	J^{R,1}_\theta
	&=&
	\limsup_{h\to+0}
	\limsup_{n\to\infty}
	\frac
	{
	( 1+ h^2 J^R_{\theta,h/\sqrt n} /n )^n
	-1
	}
	{h^2}
	=
	\limsup_{h\to+0}
	\frac
	{
	e^{h^2 J^R_\theta}
	-1
	}
	{h^2}
	\nonumber\\
	&=&
	J^R_\theta
	.
	\label{eq:jrtheta}
\end{eqnarray}
(The left differentiable case is also given in a similar way.)

Next,
we define asymptotic unbiasedness for a sequence
$M_n$ of estimators of $g(\theta)$ based on measurements of
$\rho_\theta^{\otimes n}$ as follows.

\begin{definition}
\rm
A sequence $M_n$ of  estimators of $g(\theta)$ is said to be
$c_n$-unbiased at $\theta$ from the right if
\[
	c_n b(M_n,\theta+h/c_n)\to0 \ \mbox{ as }n\to\infty
\]
for any $h\ge0$ which is small enough.
Similarly,
$M_n$ is said to be
$c_n$ unbiased at $\theta$ from the left if
\[
	c_n b(M_n,\theta-h/c_n)\to0 \mbox{ as }n\to\infty
\]
for any $h\ge0$ small enough.
\end{definition}

Then, we have the following
asymptotic right (resp. left) QHCRK inequality.

\begin{theorem}\label{th:AQHCRC}
Assume that
$M_n$ is $c_n$ unbiased at $\theta$ from the right 
(resp. from the left),
and $J^{R,1}_\theta<\infty$,
$|g'_1(\theta)|<\infty$
( resp. 
$J^{R,0}_\theta<\infty$, 
$|g'_{0}(\theta)|<\infty$).
Then,
\begin{equation}
	\label{eq:aQHCR}
	\liminf_{n\to\infty}c_n^2 V_\theta(M_n)
	\ge(g'_1(\theta))^2/J^{R,1}_\theta
\end{equation}
(resp.
\[
	\liminf_{n\to\infty}c_n^2 V_\theta(M_n)
	\ge(g'_{0}(\theta))^2J^{R,0}_\theta
	\quad \mbox{\it )}
\]

\end{theorem}

The proof is given in Section \ref{sec:proof}.

\begin{remark}
\rm 
 Due to (\ref{eq:jrtheta}), theorem~\ref{th:AQHCRC} implies
AQCR inequality. Notice that this argument is technically much more complicated
than the proof of the similar statement in non-asymptotic theory.
In fact, it is not straightforward to prove
a similar relation between asymptotic SLD-base inequalities, 
for the absence of the equivalence of the equation~(\ref{eq:3:tohi}).
\end{remark}

\subsection{Non-asymptotic theory for a model with a vector valued parameter}
Define an operator $\Delta_{i,\delta}^{t}$ by,
\[
	\Delta_{i,\delta}^{t} f(\theta)
	=
	\frac{f(\theta^1,...,\theta^i+t\delta,...,\theta^m)
     -f(\theta^1,...,\theta^i-(1-t)\delta,...,\theta^m)}{\delta},
\]
and difference version of SLD $L^{S,t}_{i,\theta,\delta}$
and RLD $L^{R,t}_{i,\theta,\delta}$
is defined as 
the solution to the equations
\begin{align*}
\Delta_{i,\delta}^{t}\rho_{\theta}
=\frac{1}{2}(L^{S,t}_{i,\theta,\delta}\rho_\theta 
        + \rho_\theta L^{S,t}_{i,\theta,\delta})
=\rho_\theta L^{R,t}_{i,\theta,\delta}.
\end{align*}

The following relation will be useful for the further analysis.

For ${\bf t}=(t_i,...,t_m)$ $(0\leq t_i\leq 1)$ and
$\mbi{\delta}=(\delta_1,...,\delta_m)$,
define matrices 
$J^{S,{\bf t}}_{\theta,\mbi{\delta}}$, $J^{R,{\bf t}}_{\theta,\mbi{\delta}}$ by,
\begin{align*}
J^{S,{\bf t}}_{\theta,\mbi{\delta},i,j}
=
\frac{1}{2}
\tr\rho_\theta
 (L^{S,t_i}_{i,\theta,\delta_i}L^{S,t_j}_{j,\theta,\delta_j}+
L^{S,t_j}_{j,\theta,\delta_j}L^{S,t_i}_{i,\theta,\delta_j}
),
\quad\quad\quad
J^{R,{\bf t}}_{\theta, \mbi{\delta},i,j}
=
\tr\rho_\theta
L^{R,t_j}_{j,\theta,\delta_j}(L^{R,t_i}_{i,\theta,\delta_j})^\dagger,
\end{align*}

Analogically to models with scalar parameters,
we have the following theorem, which is proven in Section~\ref{sec:proof}.
\begin{theorem}
\label{theorem:qhcrk-multi}
If the estimator is unbiased, for any $\mbi{\delta}$ and for any ${\bf t}$
such that  $0\leq t_i\leq 1$, we have,
\begin{align*}
{\rm Sp} G V_\theta(M)
&\geq 
{\rm Sp} G(J^{S,{\bf t}}_{\theta,\mbi{\delta}})^{-1},\\
  {\rm Sp} G V_\theta(M)
&\geq 
{\rm Sp} G(J^{R,{\bf t}}_{\theta,\mbi{\delta}})^{-1}
+{\rm Spabs}\Im G(J^{R,{\bf t}}_{\theta,\mbi{\delta}})^{-1}.
\end{align*}
\end{theorem}

\section{Examples}
\label{sec:example}
In this section, we will demonstrate our theory in 
several examples.

\subsection{Estimation of concurrence}\label{sec:conc}

For $-1<\theta<1$,
let $\rho_\theta$ be a density matrix on
a $2\times2$ system given by
\[
	\rho_\theta=
	\frac{1+\theta}2\ket{\Phi^+}\bra{\Phi^+}
	+
	\frac{1-\theta}2\ket{\Phi^-}\bra{\Phi^-}
\]
where $\ket{\Phi^\pm}=(\ket{00}\pm\ket{11})/\sqrt2$
are maximally entangled states which are mutually orthogonal.
This type of state can be produced
by parametric down conversion (\cite{kwiat}).

Let $g(\theta)=|\theta|$, which is equal to
the concurrence (\cite{wootters}),
a measure of the quantum entanglement
between two 2-level systems.
At $\theta\ne0$, we can define the usual SLD and RLD,
and both $J^S_\theta$ and $J^R_\theta$ are given by
\[
	J^S_\theta=J^R_\theta=\frac1{1-\theta^2}.
\]
At $\theta=0$, we have
\[
	J^{S,1}_{\theta,\delta}=J^{R,1}_{\theta,\delta}=
	\frac1{1-\theta^2}
\]
for any $-1<\delta<1$.
Therefore, QHCRK inequality is
\[
	V_\theta(M)\ge1-\theta^2.
\]
In this model, however,
there is no unbiased estimator for $g(\theta)=|\theta|$.
Therefore,
the inequality is nonsense in the non-asymptotic setting.

Let us consider the asymptotic case,
where we construct a sequence $M_n$ of estimators
of $g(\theta)$, which achieves the asymptotic AQHCRK bound.
Here, $c_n=\sqrt n$.
The AQHCRK inequality for estimators
which are $\sqrt n$-unbiased from the right
is given due to (\ref{eq:jrtheta})
\[
	\lim_{n\to\infty}c_n^2 V_\theta(M_n)\ge
	1-\theta^2.
\]
This bound is achieved by the following scheme.
\\
{\bf Step 1} \
For $i=1,2....,n$,
we measure $i$-th subsystem of $\cl H^{\otimes n}$
by a two valued POVM
$\{\ket{\Phi^+}\bra{\Phi^+},I-\ket{\Phi^+}\bra{\Phi^+}\}$.
Let $x_i=1$ if $\ket{\Phi^+}\bra{\Phi^+}$ is observed, and 
let $x_i=0$ if $I-\ket{\Phi^+}\bra{\Phi^+}$ is observed.
\\
{\bf Step 2} \
Let
\[
	y=
	\begin{cases}
	\sum_{i=1}^n x_i/n
	 & $if $ \sum_{i=1}^n x_i/n \ge -n^{-1/3},\cr
	-\sum_{i=1}^n x_i/n
	 & $if $ \sum_{i=1}^n x_i/n  <  -n^{-1/3}.
	\end{cases}
\]
The expectation $E_\theta(y)$ of $y$ for $\rho_\theta^{\otimes n}$ is
$\theta+o(1/n)$ as $n\to\infty$ so such an estimator $M_n$ is
$1/\sqrt n$-unbiased at $\theta$ from the right,
and $V_\theta(M_n)=(1-\theta^2)/n$.

\subsection{A discrete model}\label{sec:discrete}

This example can be regarded as a non-commutative version of
the classical model of discrete uniform distributions.

Define $2\times2$ matrices $\sigma_1$, $\sigma_2$ and $0_2$ as
\[
	\sigma_1=\begin{pmatrix}1&0\cr0&0\end{pmatrix},\
	\sigma_2=\begin{pmatrix}1&1/2\cr1/2&1\end{pmatrix},\
	0_2=\begin{pmatrix}0&0\cr0&0\end{pmatrix}.
\]
Let $\cl H$ be an infinite-dimensional Hilbert space
spanned by $\ket1,\ket2,\ket3,...$.
For $\theta\in\Theta$($=\mathbb N$: the set of natural numbers),
let
$\rho_\theta\in\cl S(\cl H)$ be a density operator of the form
\[
	\rho_\theta=
	\begin{cases}
	\theta^{-1}{\rm diag}
	(\underbrace{\sigma_2,...,\sigma_2}_{(\theta-1)/2},\sigma_1,
	0_2,0_2,...)
	& $if $\theta$ is odd,$ \cr
	\theta^{-1}{\rm diag}
	(\underbrace{\sigma_2,...,\sigma_2}_{\theta/2},0_2,0_2,...)
	& $if $\theta$ is even$
	\end{cases}
\]
where ${\rm diag}(...)$  means the diagonal matrix.

\begin{remark}\label{rem:span}
\rm
Though $\rho_\theta$ is not invertible on $\cl H$,
the assumption (\ref{ass:inverse}) essentially applies to
this example.
Indeed,
for the achievability of the lower bound,
it is sufficient to consider
the subspace $\cl H'_\theta(\subset\cl H)$
supported by $\rho_\theta$.
\end{remark}

\subsubsection*{One-sample analysis}

When $\theta$ is even,
there is an estimator of $\theta$ which minimizes variance.
Define a hermitian matrix $T$ by
\[
	T=
	{\rm diag}(T_1,T_3,T_5,...)
\]
where, for an odd number $i$,
\[
	T_i=\begin{pmatrix}
		2i-1 & 1/2 \cr 1/2 & 2i+1/2
	\end{pmatrix}.
\]
Let the spectral decomposition, described by $M$, be the estimator of $\theta$.
Since
\[
	\tr(\rho_\theta T) =\theta
	\
	\mbox{ for any }\theta,
\]
$M$ is an unbiased estimator.

If $\theta$ is even, then $M$ is optimum.
The proof is as follows.
Define
$\Lambda_0,\Lambda_1,...,\Lambda_{\theta-1}$
by
\[
	\rho_k=
	\frac
	{\rho_\theta\Lambda_k+\Lambda_k\rho_\theta}
	{2}
	\
	(k=0,1,...,\theta-1)
	,
\]
that is,
\[
	\Lambda_i=
	\begin{cases}
	\displaystyle
	\frac{\theta}{\theta-i}{\rm diag}
	(\underbrace{1,...,1}_{\theta-i},0,0,...)
	& \mbox{if }i\mbox{ is even,}
	\cr
	\displaystyle
	\frac{\theta}{\theta-i}{\rm diag}
	\Big(\underbrace{1,...,1}_{\theta-i-2},
	\begin{pmatrix}
		-1/6 & 1/3 \cr 1/3 & 5/6
	\end{pmatrix}
	,0,0,...\Big)
	& \mbox{if }i\mbox{ is odd.}
	\end{cases}
\]
Since
\begin{eqnarray*}
	\Delta_{-1,k}\rho_\theta
	&=&
	\frac
	{
	\rho_\theta L^S_{\theta,-1,k}
	+
	L^S_{\theta,-1,k} \rho_\theta
	}
	{2}
	\\
	&=&
	\frac
	{
	\rho_\theta
	\big((-1)^k\sum_{i=0}^k(-1)^i{k\choose i}
	\Lambda_i\big)
	+
	\big((-1)^k\sum_{i=0}^k(-1)^i{k\choose i}
	\Lambda_i\big)
	\rho_\theta
	}
	{2}
	,
\end{eqnarray*}
we have
\[
	L^S_{\theta,-1,k}
	=
	(-1)^k
	\sum_{i=0}^k
	(-1)^i
	{k\choose i}
	\Lambda_i
	.
\]
That is, there is a triangle matrix $U$
with non-zero diagonal elements
such that
$(L^S_{\theta,-1,0},L^S_{\theta,-1,1},...,
L^S_{\theta,-1,\theta-1})
=
U
\tran(\Lambda_0,\Lambda_1,...,\Lambda_{\theta-1})$.
This means that
$\{L^S_{\theta,-1,k}\}$ and $\{\Lambda_k\}$
have a one-to-one correspondence.
Since
$T$ is
an element of
${\rm span}\{\Lambda_0,\Lambda_1,...,
\Lambda_{\theta-1}\}$,
$T-\theta\in{\rm span}
\{
L^S_{\theta,-1,1},...,L^S_{\theta,-1,\theta-1}
\}$.
Hence the equality in (\ref{eq:th2:s}) holds.

By direct calculation,
the MSE of $M$ is given by
\[
	V_\theta(M)=
	\begin{cases}
		\theta^2/3-7/12+1/(2\theta) & $if $\theta$ is odd,$\cr
		\theta^2/3-7/12 & $if $\theta$ is even.$
	\end{cases}
\]

When $\theta$ is odd,
a similar argument of $L^S_{\theta,-1,k}$ implies that
the best estimator should be given by the spectral decomposition of
$T'$ of the form
\[
	T'=
	{\rm diag}
	(T_1,T_3,...,T_{\theta-2},T'_\theta,T_{\theta+2,...})
\]
where
\[
	T'_\theta=\begin{pmatrix}
		2\theta-1 & 0 \cr 0 & 2\theta+1
	\end{pmatrix}.
\]
The MSE of the estimator is
\[
	\theta^2/3-7/12+1/(4\theta)
\]
for odd $\theta$.
However, $T'$ depends on the unknown parameter
$\theta$.
In this sense, there is no estimator which is uniformly
achieves the QK bound.

\subsubsection*{Asymptotic analysis}

Let us consider the asymptotic setting.
$\rho_\theta^n=\rho_\theta^{\otimes n}$.
Since this model is discrete and
each element $\rho_\theta$ is isolated, due to 
Theorem \ref{th:AQHCRD}, we have the following lower bound
to $\liminf_{n\to\infty}\log V_{\theta}(M_n)$,
\[
      \log (1+ J^{R,1}_{\theta,-1})=
	\begin{cases}
	\log(\theta/(\theta-1)) & $if $\theta$ is odd$,\cr
	\log(\theta/(\theta-1))+\log\frac{3\theta-2}{3(\theta-1)}
	 & $if $\theta$ is even$.
	\end{cases}
\]
Based on asymptotically optimal test between  two hypothesis,
we will construct an estimator $M$,
which achieves the AQHCRK bound to the discrete case of Theorem \ref{th:AQHCRD}
if $\theta$ is odd.

$M$ is constructed as follows.
First,
let $\Theta'$ be the set of positive even numbers.
Next,
let $M'$ be a POVM of the form
\[
	M(\theta')=
	{\rm diag}(\underbrace{0,...,0}_{\theta'-2},I_2,0,0,...)
\]
where $I_2$ is the $2\times2$ identity matrix.
We independently apply $M'$
to each sample,
and let $\hat\theta'_i\in\Theta'$ be
the observed value for the $i$-th sample.
Let
$\hat\theta_{\rm max}=
	\max\{\hat\theta'_1,...,\hat\theta'_n\}$.
Our next step is estimation in a submodel
$\Theta'=\{\hat\theta_{\rm max},\hat\theta_{\rm max}-1\}\subset\Theta$.
Let $N_n$ be a POVM given in (\ref{eq:hiaipetz}) where
$\theta=\hat\theta_{\rm \max}-1$ and $\theta+\delta=\hat\theta_{\rm \max}$.
Since each $M'(\theta')$ commutes with
$\rho_{\hat\theta_{\rm \max}-1}$ and $\rho_{\hat\theta_{\rm \max}}$,
due to \cite{hiaipetz}, $M'$ and $N_n$ commute.
Applying such a measurement $N_n$,
the estimate $\hat\theta$ of $\theta$ is given by
\[
	\hat\theta=
	\begin{cases}
	\hat\theta_{\rm \max}-1
	& $if $ \hat\theta_{\rm \max}-1\in\Theta'$ is observed,$\cr
	\hat\theta_{\rm \max}
	& $if $ \hat\theta_{\rm \max}\in\Theta'$ is observed.$
	\end{cases}
\]
The exponent of MSE of this estimator $M$ is given by
\[
	\liminf_{n\to\infty}\frac1n\log V_\theta(M)=
	\begin{cases}
	-D(\rho_{\theta-1}\parallel\rho_\theta)
         =
	- \log (1+J^{R,1}_{\theta,-1})
	 & $if $\theta$ is odd,$\cr
	-D(\rho_{\theta-1}\parallel\rho_\theta)
         >
        - \log (1+J^{R,1}_{\theta,-1})
	 & $if $\theta$ is even$
	\end{cases}
\]
where
\[
	D(\rho_{\theta-1}\parallel\rho_\theta)=
	\log\frac{\theta}{\theta-1}
	+\frac1{\theta-1}\log\frac{2}{\sqrt3}
\]
if $\theta$ is even.

\subsection{Submodels of Gaussian state model}
We consider singular submodels of Gaussian state model,
whose theory (\cite{yuenlax, holevo}) is reviewed in the next
subsection.
In last subsection of this section,
we study a vector valued submodel of Gaussian state model,
to see the effect of non-commutativity explicitly.
Before that, for preparation, we study
a submodel of this model, which has scalar valued parameter.
\subsubsection*{Gaussian state model}
For simplicity, we chose the unit such that $\hbar=1$.
Gaussian state model is a family of states living in the Fock space,
${\cal H}={\rm span}\{\ket{n};n=0,1,2,\cdots\}$,
such that 
\begin{align*}
\rho^G_\theta=
\int\frac{{\rm d}p{\rm d}q}{2\pi(\sigma^2-1/2)}
e^{-\frac{(p-\theta^1)^2+(q-\theta^2)^2}{2(\sigma^2-1/2)}}\ket{p,q}\bra{p,q},
\end{align*}
where $\ket{p,q}$ is a coherent state.
After some calculations, one would obtain (\cite{holevo}),
\begin{align}
&L^S_{1,\theta}=\frac{1}{\sigma^2}(P-\theta^1),\quad
 L^S_{2,\theta}=\frac{1}{\sigma^2}(Q-\theta^2),\nonumber\\
& L^R_{1,\theta}
=\frac{1}{\sigma^4-1/4}(\sigma^2(P-\theta^1)+\frac{\sqrt{-1}}{2}(Q-\theta^2)),
\quad
 L^R_{2,\theta}
=\frac{1}{\sigma^4-1/4}(\sigma^2(Q-\theta^2)-\frac{\sqrt{-1}}{2}(P-\theta^1)),
\nonumber\\
&J^S_\theta=\frac{1}{\sigma^2}\left[
\begin{array}{cc}
 1&0 \\
 0&1
\end{array}
\right],\quad
J^R_\theta=\frac{1}{\sigma^4-1/4}\left[
\begin{array}{cc}
 \sigma^2&-\frac{\sqrt{-1}}{2} \\
 \frac{\sqrt{-1}}{2}&\sigma^2
\end{array}
\right].
\label{eq:gaussian:sldrld}
\end{align}

The submodel $\{\rho^G_{(\theta^1_0,\theta^2)};\theta^2\in\mathbb{R}\}$
and the submodel
$\{\rho^G_{(\theta^1,\theta^2_0)};\theta^1\in\mathbb{R}\}$
are exponential families, and 
the equality in the SLD QCR inequality,
\begin{align*}
 V_\theta(M)\geq \sigma^2.
\label{eq:bound3}
\end{align*}
 is achieved for both models, 
the former by the spectral decomposition of $P$, and 
the latter by the spectral decomposition of $Q$.

For the model $\{\rho^G_\theta;\theta\in\mathbb{R}^2\}$ 
with 2-dim parameter, it is known that the equality in the RLD QCR
inequality is achieved. Namely, with  $G=I$, the RLD  and 
the SLD QCR inequality is given by
\begin{align*}
 {\rm Sp}V_{\theta}(M)&\geq 2\sigma^2+1,\\
 {\rm Sp}V_{\theta}(M)&\geq 2\sigma^2,
\end{align*}
respectively.
For SLD's, being scaler multiplications of  $P$ and $Q$,
are not commutative,
the achievable bound is larger
than the bound by SLD QCR.

If the thermal noise is very large and thus $\sigma$ is very large,
the difference between SLD QCR and RLD QCR is negligible,
meaning that the effect of non-commutativity is relatively negligible.

Letting $\ket{p,q}$ be a coherent state, 
the optimal estimator is 
\[
\{\ket{\theta^1,\theta^2}\bra{\theta^1,\theta^2}\}. 
\]
This POVM is often referred to as optimal 
simultaneous measurement of $P$ and $Q$,
for this satisfies the relation
\begin{equation}
 \int \frac{{\rm d}p{\rm d}q}{2\pi} p\ket{p,q}
 \bra{p,q}=P,\quad
 \int \frac{{\rm d}p{\rm d}q}{2\pi} q\ket{p,q}
 \bra{p,q}=Q,
\label{eq:decomPQ}
\end{equation}
and 
gives the smallest sum of MSE's 
in all such POVM's.

\subsubsection*{A singular submodel with scalar parameter}
The model $\{\rho_\theta;\theta\in\mathbb{R}\}$, where
\begin{align*}
 \rho_\theta=\left\{
   \begin{array}{cc}
    \rho^G_{(0,0)}\otimes\rho^G_{(\theta,0)}&  (\theta\leq 0)\\
    \rho^G_{(\theta,0)}\otimes\rho^G_{(0,0)}&  (\theta\geq 0)\\
   \end{array}
   \right. ,
\end{align*}
has a singularity at $\theta=0$, and the singular point is
differentiable both from the left and the right.
In the following, we denote by $P_i$ and $Q_i$ 
the quadrature operators of the mode $i$.

SLD QHCRK inequality for $\theta\neq 0$ coincide with
SLD QCR inequality, and is given by,
\begin{align}
 V_{\theta}(M)\geq 
   \sigma^2 \quad\quad(\theta\neq 0).
\label{eq:bound2}
\end{align}
The equality is achieved by the spectral decomposition of 
$P_2$ (resp. $P_1$) if $\theta<0$ (resp. $\theta>0$), in this region of
the parameter.
However, neither $P_1$ nor $P_2$ is unbiased all over the model.

At the point $\theta=0$, due to (\ref{eq:16}) and (\ref{eq:gaussian:sldrld}), we have
\begin{align*}
 \lim_{\delta\to 0+}L^{S,t}_{\theta, \delta}=
\frac{1}{\sigma^2}(t P_1+ (1-t)P_2),
\end{align*}
and letting $t=\frac{1}{2}$, we obtain the lower bound,
\begin{align}
 V_{0}(M)\geq 2\sigma^2.
\label{eq:bound1}
\end{align}
The estimator given by the spectral decomposition of $P_1+P_2$
is unbiased at all $\theta\in\Theta$, and achieves the equality in 
the inequality (\ref{eq:bound1}). This estimator,
however, does not achieve the lower bound by (\ref{eq:bound2})
at $\theta\ne0$.

We conjecture,
however, this estimator is optimal,
and that the lower bound~(\ref{eq:bound1}) is a lower bound
all over the model.
Indeed, we can improve the lower bound~(\ref{eq:bound2})
as follows. Almost in parallel with Theorem~\ref{th:QK},
we have the following Koike type inequality:
\begin{align}
 V_\theta(M)\geq (1\;\; 1)
  \left(\begin{array}{cc}
     \tr\rho_\theta L^{R,1}_{\theta,\delta_2}
   (L^{R,1}_{\theta,\delta_1})^\dagger  &    
   \tr\rho_\theta L^{R,1}_{\theta,\delta_2}
   (L^{R,1}_{\delta_2,\theta})^\dagger \\
   \tr\rho_\theta L^{R,1}_{\theta,\delta_2}
   (L^{R,1}_{\delta_1,\theta})^\dagger &
     \tr\rho_\theta L^{R,1}_{\theta,\delta_1}
   (L^{R,1}_{\theta,\delta_1})^\dagger      
	 \end{array}\right)^{-1}
   \left(\begin{array}{c}
     1 \\
     1
	 \end{array}\right).
\end{align}
Let us chose infinitesimally small $\delta_2$, or
take the limit $\delta_2\to 0.$
Then, the right hand side is,
\begin{align}
\frac{1}{J}
+\frac{(\delta_1)^2}
{e^{J((\theta)^2+(\delta_1)^2)} -1-(\theta)^2J},
\label{eq:bound-ex-gauss-1}
\end{align}
with $J$'s being 
the $(1,1)$-component of $J^R_\theta$ for
the model $\{\rho^G_{\theta};\theta\in\mathbb{R}^2\}$,
which is given in the equations~(\ref{eq:gaussian:sldrld}).

The detail of the calculation is omitted,
but the key is the following equality:
\begin{align}
 \tr\rho^{G-1}_0\rho^G_{(x,y)}\rho^G_{(z,w)}=
e^{J(xz+yw)-A(xw-yz)},
\label{eq:trace}
\end{align}
with $A$ being the $(1,2)$-component of $J^R_\theta$ for
the model $\{\rho^G_{\theta};\theta\in\mathbb{R}^2\}$,
which is given in the equations~(\ref{eq:gaussian:sldrld}).
The rough derivation of the equation~(\ref{eq:trace}) will be
given in Section~\ref{sec:proof}.

\begin{figure}
\hfil\includegraphics[scale=0.7]{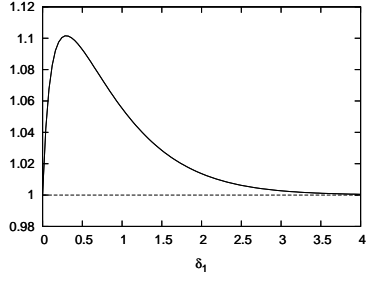}
\caption{\label{gauss-bound.eps} A Koike-type lower bound is  plotted against 
$\delta_1$,
when $\sigma=1$ and $\theta=1$. 
With optimal $\delta_1$, the new equality improves
SLD QHCRK inequality with $\delta\to 0$, 
which equals $1\; (=\sigma^2)$.
}
\end{figure}

Let us  chose $\delta_1$ to be very small,
or take the limit $\delta_1\to 0$,
and consider the case where 
$\frac{(\theta)^2}{\sigma^2}\ll 1$ and 
$\frac{1}{\sigma^2}\ll 1$ .
Then, the bound (\ref{eq:bound-ex-gauss-1})
is very close to the bound~(\ref{eq:bound1}), which we conjecture is
the tight bound all over the model.
At least, as in Fig.~\ref{gauss-bound.eps}, 
the new lower bound improves the bound by SLD QHCRK inequality with
$\delta\to 0$ in some cases.

\subsubsection*{A singular model with the vector valued parameter}
The model
$\{\rho_\theta;\theta\in\mathbb{R}\}$, where
\begin{align*}
 \rho_\theta=\left\{
   \begin{array}{cc}
    \rho^G_{(0,0)}\otimes\rho^G_{(\theta^1,\theta^2)}
     &  (\theta^1\leq 0,\theta^2\leq 0)\\
    \rho^G_{(\theta^1,0)}\otimes\rho^G_{(0,\theta^2)}
     &  (\theta^1\geq 0,\theta^2\leq 0)\\
    \rho^G_{(0,\theta^2)}\otimes\rho^G_{(\theta^1,0)}
     &  (\theta^1\leq 0,\theta^2\geq 0)\\
    \rho^G_{(\theta^1,\theta^2)}\otimes\rho^G_{(0,0)}
     &  (\theta^1\geq 0,\theta^2\geq 0)\\
   \end{array}
   \right. ,
\end{align*}
has singular point at $\theta=0$,
and the singular point is
differentiable both from the left and the right.
The parameter space $\Theta$ is broken down into 
the following three parts according to types of singularity.
\begin{itemize}
 \item[(1)] $\theta^1\neq 0$ and $\theta^2\neq 0$: differentiable
 \item[(2)] $\theta^1\neq 0$ or $\theta^2\neq 0$: differentiable 
            with respect to $\theta^1$ or $\theta^2$.
 \item[(3)]  $\theta^1= 0$ and $\theta^2= 0$:  un-differentiable 
             with respect to  $\theta^1$ and $\theta^2$.
\end{itemize}

In case of  (1), 
QHCRK inequalities for $\theta\neq 0$ coincide with
their differentiable versions (QCR's).
Namely, in the region  
$\{\theta ;\theta^1 \leq 0, \theta^2 \leq 0\}$
the bound by the RLD QHCRK inequality
\begin{align}
 {\rm Sp} V_{\theta}(M)\geq 
   2\sigma^2+1 
\label{eq:gauss-singular-bound-1}
\end{align}
is achieved by the estimator
$\{\ket{\theta^1,\theta^2}\bra{\theta^1,\theta^2}\otimes I\}$,
where $\ket{\theta^1,\theta^2}$ is a coherent state.
The similar holds for 
the region
$\{\theta ; \theta^1 \geq 0, \theta^2 \geq 0\}$.
In the region $\{\theta ; \theta^1 \leq 0, \theta^2 \geq 0\}$ 
the bound by the SLD QHCRK inequality
\begin{align*}
 {\rm Sp} V_{\theta}(M)\geq 
   2\sigma^2 
\end{align*}
is achieved by the simultaneous spectral decomposition of 
$P_2\otimes Q_1$.
The similar holds for 
the region $\{\theta ; \theta^1 \geq 0, \theta^2 \geq 0\}$.
However, none of those 'locally optimal' estimators is 
unbiased all over the model.

Next, we consider case of (2).
Namely, we consider the region 
$\{\theta ; \theta^1\leq 0, \theta^2=0\}$.
(Other regions in this case are essentially the same.)
If $\theta^1\leq 0$ and $\theta^2=0$,
due to (\ref{eq:16}) and (\ref{eq:gaussian:sldrld}),
we can compute 
$ \lim_{\delta\to 0+}L^{S,t_i}_{i, \theta, \delta}\, $
$ \lim_{\delta\to 0+}L^{R,t_i}_{i, \theta, \delta}\, (i=1,2)$,
explicitly,
and the RLD and the SLD QHCRK inequality is obtained as,
\begin{align*}
{\rm Sp}V_\theta(M)&\geq 
\frac{(\sigma^4-\frac{1}{4})}{\sigma^4(2t_2^2-2t_2+1)-\frac{(1-t_2)^2}{4}}\{2\sigma^2(t_2^2-t_2+1)+1-t_2\},\\
{\rm Sp}V_\theta(M)&\geq 
\frac{\sigma^2}{2t_2^2-2t_2+1}+\sigma^2,
\end{align*}
respectively.

Now, our concern is the value of $t_2$ which gives the largest lower bound.
If $\sigma$ is very large (thus effect of non-commutativity is
negligible), then the right hand side of  the RLD QHCRK is almost
the same as the right hand side of the SLD QHCRK, 
and gives the largest lower bound at $t_2=1/2$.  
On the other hand, 
let us consider the opposite extreme case where
 the noise is purely quantum, that is, the Gaussian states 
are coherent states. This case is obtained by letting $\sigma^2=1/2$.
If and only if $t_2=0$,
the right hand side of the RLD QHCRK 
gives the non-trivial lower bound,
\begin{align*}
{\rm Sp}V_\theta(M)&\geq 2.
\end{align*}
(If $t_2\neq 0$, the right hand side of the inequality vanishes.)

To sum up,  the value of $t_2$ which gives best bound is dependent on the 
non-commutativity inherent in the model. 

\begin{figure}
\hfil\includegraphics{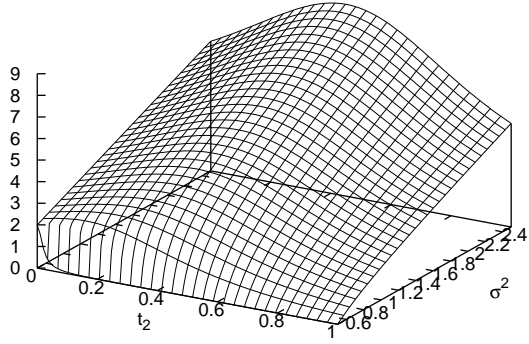}
\caption{\label{RLD-QCHK} The lower bound by the QHCRK inequality is
 plotted against $t_2$ and $\sigma^2$. 
When $\sigma=\frac{1}{2}$ (no thermal noise), 
the largest lower bond is given by $t_2=0$.
As $\sigma^2$ tends to larger, the peak  moves towards $t_2=\frac{1}{2}$.
}
\end{figure}

Finally, the case of (3) is analyzed. 
Here, we have 
Due to (\ref{eq:16}) and (\ref{eq:gaussian:sldrld}),
$\lim_{\delta\to 0+}L^{R,t_i}_{i, \theta, \delta}\,(i=1,2)$
are computed explicitly.
Letting $t_1=t_2=1/2$, we have the following RLD QHCRK inequality,
\begin{align}
 {\rm Sp}V_\theta(M)\geq 4\sigma^2+2.
\label{eq:bound4}
\end{align}
This choice of the parameters $t_1,t_2$ is optimal, for
the unbiased estimator which is constructed below 
achieves this lower bound.

Let 
$
 P_{1^*}=\frac{P_1+P_2}{\sqrt{2}},\quad
 Q_{1^*}=\frac{Q_1+Q_2}{\sqrt{2}},\quad
 P_{2^*}=\frac{P_1-P_2}{\sqrt{2}},\quad
 Q_{2^*}=\frac{Q_1-Q_2}{\sqrt{2}}.
$
Then, $ P_{i^*},Q_{i^*}$ 
satisfies the canonical commutation relation,
and $P_{i^*}$ , $Q_{j^*}$ $(i\neq j)$ commutes.
Consider the new split of the system into 
the mode $1^*$ and $2^*$, introduce 
a new coherent state $\ket{p,q}^*$
with respect to $P_{1^*}$ and $Q_{1^*}$,
and define an estimator
$$
 \left\{
  \ket{\theta^1/\sqrt{2},\theta^2/\sqrt{2}}^*\;
  \,^*\bra{\theta^1/\sqrt{2},\theta^2/\sqrt{2}}
 \otimes I
\right\}.
$$
Observing that the expectation of $P_1+P_2$ and  $Q_1+Q_2$
coincides with $\theta^1$ and  $\theta^2$ respectively,
unbiasedness of the estimator is checked by using (\ref{eq:decomPQ}).
A straight forward calculation shows that this estimator
achieves the lower bound implied by (\ref{eq:bound4}).

This estimator is optimal at $\theta=0$, and unbiased all over the
model. We conjecture that 
the lower bound at other points of the model is improved to show
the optimality of this estimator.

In fact, for example, the lower bound
(\ref{eq:gauss-singular-bound-1}) 
is improved 
by using RLD QHCRK inequality with a finite $\delta$.
Let $\theta^1<0,\;\theta^2<0$, and 
$\delta_1=-\theta^1+t,\;\delta_2=-\theta^2+s$, for $s>0,t>0$.
Then, due to (\ref{eq:trace}), the RLD QHCRK inequality is calculated as,
\begin{align*}
{\rm Sp}V_{\theta}(M)\geq \frac{a+d+2|c|}{ad-b^2-c^2}.
\end{align*}
Here, letting $J$ and $A$ be the $(1,1)$- and the $(1,2)$- component
of the $J^R_\theta$ for the model
$\{\rho^G_\theta;\theta\in\mathbb{R}^2\}$, respectively, 
\begin{align*}
 a&=\frac{e^{J((\theta^1)^2+(t)^2)}-1}{(\theta^1+t)^2},
\quad  
d=\frac{e^{J((\theta^2)^2+(s)^2)}-1}{(\theta^2+s)^2},\quad
 b+\sqrt{-1}c= e^{-A(ts+\theta^1\theta^2))}-1.
\end{align*}
When $\frac{1}{\sigma^2}\ll 1$ 
and $\frac{(\theta^i)^2}{\sigma^2}\ll 1\; (i=1,2)$,
letting $t=\theta^1$ and $s=\theta^2$,
the bound is nearly $4\sigma^2$,
and is close to (\ref{eq:bound4}),
which is conjectured to be a lower bound all over the model.

\begin{figure}
\hfil\includegraphics{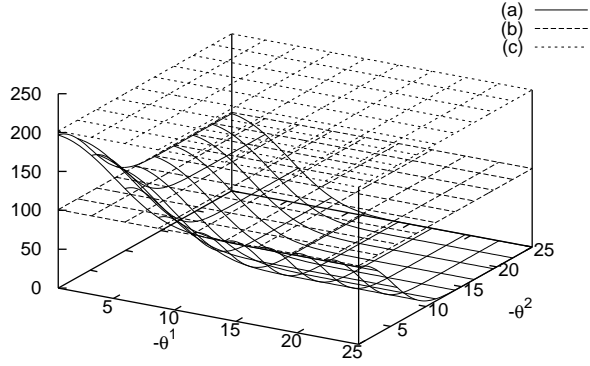}
\caption{\label{fig:gauss-singular-2dim} 
 RLD QHCRK bound 
with $\delta^i=2\theta^i$ ((a)),
with $\delta^i\to 0$ ((b)) are  plotted against $\theta\; (\theta^i<0)$.
(c) is the conjectured lower bound (and is RLD QHCRK bound at
 $\theta=0$). (a) is above (b) in some region,
 and is very close to (c) around the origin.
}
\end{figure}

As is indicated in Fig.~\ref{fig:gauss-singular-2dim},
we can observe in some  region of the parameter space,
the new bound, RLD QHCRK bound with $\delta^i=2\theta^i$ 
\;$(\theta^i<0)$
improves (\ref{eq:gauss-singular-bound-1}).

\section{Proofs}
\label{sec:proof}
In this section, we give proofs of theorems.

First, we prepare a lemma which says that the best unbiased estimator is
a projection valued measure (PVM).
\begin{lemma}
Let
\[
	T=\int_{\gamma\in g(\Theta)}\gamma M(d\gamma).
\]
Then, it holds that
\[
	\int_{\gamma\in g(\Theta)}\gamma^2 \tr(\rho_\theta M(d\gamma))
	\ge
	\tr(\rho_\theta T^2).
\]
\end{lemma}
{\bf Proof.} \
\[
	\int_{\gamma\in g(\Theta)}
	\gamma^2 \tr(\rho_\theta M(d\gamma))
	-
	\tr(\rho_\theta T^2)
	=
	\tr\Big(\rho_\theta
	\int_{\gamma\in g(\Theta)}
	(\gamma-T)^2
	M(d\gamma)\Big)
	\ge0.
\]
\hfill$\Box$

Hence we use an observable $T$ to describe the POVM $M$ in Proofs of
Theorems \ref{th:QHCR} and \ref{th:QK}.

\subsection{Proof of Theorem \ref{th:QHCR}}
Let $T\in\cl L(\cl H)$ be an operator satisfying
$\tr(\rho_\theta T)=g(\theta)$.
By Schwartz's inequality,
it holds that
\begin{eqnarray}
	&&
	\tr(\rho_\theta(T-g(\theta))^2)
	\tr(\rho_\theta (L^S_\theta)^2)
	\ge
	\Big(
	\tr\big(
	\rho_\theta(T-g(\theta) L^S_\theta
	\big)
	\Big)^2
	\nonumber
	\\
	&&
	=
	\Big(
	\tr\big(
	\big(T-g(\theta)\big)
	(\Delta_\delta\rho_\theta)
	\big)
	\Big)^2
	=
	(\Delta_{\delta}g(\theta))^2.
	\label{eq:QHCRproof}
\end{eqnarray}
Therefore,
(\ref{eq:qhcrequality}) holds.
In addition,
\begin{eqnarray*}
	0
	&\le&
	\tr\big(
	\rho_\theta(L^S_{\theta,\delta}-L^R_{\theta,\delta})
	(L^S_{\theta,\delta}-L^R_{\theta,\delta})^\dagger
	\big)
	\\
	&=&
	\tr(\rho_\theta (L^S_{\theta,\delta})^2)
	-
	2\tr(L^S_{\theta,\delta}\Delta_\delta\rho_\theta)
	+
	\tr(\rho_\theta
	L^R_{\theta,\delta}(L^R_{\theta,\delta})^\dagger)
	\\
	&=&
	-
	\tr(\rho_\theta (L^S_{\theta,\delta})^2)
	+
	\tr(\rho_\theta
	L^R_{\theta,\delta}(L^R_{\theta,\delta})^\dagger)
	.
\end{eqnarray*}
Hence, (\ref{eq:qhcrequality-r}) holds.
The equality in (\ref{eq:QHCRproof}) holds if and only if
$T-g(\theta) = c L^S_\theta$ for a constant $c\in\mathbb R$.
\hfill$\Box$

\subsection{Proof of Theorem \ref{th:QK}}. \
Let
\[
	v=\tran
	(
	T-g(\theta),
	L^S_{\theta,\delta,1},L^S_{\theta,\delta,2},...,
	L^S_{\theta,\delta,k}
	).
\]
Let $V$ be a $k+1$-dimensional $\mathbb C$-vector space,
Then,
$v$ can be regarded as an element of $V\otimes\cl L(\cl H)$.
Let $I_V$ be the identity matrix on $V$.
Let $P=\rho_\theta\otimes I_V$.
It holds that
\[
	0\le
	\tr_{\cl H}(P v v^\dagger)
	=
	\begin{pmatrix}
	V_\theta(M)& \tran \bold v \cr
	\bold v & K^S
	\end{pmatrix}
	(=A,\mbox{ say}),
\]
where $\tr_{\cl H}$
means the partial trace over $\cl H$.
Consequently,
we obtain
\[
	V_\theta(M)
	-\tran \bold v (K^S)^{-1} \bold v
	=
	\begin{pmatrix}1& \tran \bold v (K^S)^{-1}\end{pmatrix}A
	\begin{pmatrix}1\cr (K^S)^{-1}\bold v\end{pmatrix}
	\ge0.
\]
This implies (\ref{eq:th2:s}).
By a similar argument, (\ref{eq:th2:r}) is also shown.
\hfill$\Box$

\subsection{Proof of Theorem \ref{th:AQHCRC}}
In a similar way to (\ref{eq:QHCRproof}),
Schwartz's inequality implies that
\begin{eqnarray}
	c_n^2 V_\theta(M_n)
	&\ge&
	\frac
	{(\Delta_{h/c_n} b(M_n,\theta)+\Delta_{h/c_n}g(\theta))^2}
	{\tr(\rho_{\theta}^n
	(L^{S,(n)}_{\theta,h/c_n})^2)}
	\nonumber
	\\
	\label{eq:proof:aqhcr}
	&\ge&
	\frac
	{(\Delta_{h/c_n} b(M_n,\theta)+\Delta_{h/c_n}g(\theta))^2}
	{\tr(\rho_{\theta}^n
	L^{R,(n)}_{\theta,h/c_n}(L^{R,(n)}_{\theta,h/c_n})^\dagger)}
\end{eqnarray}
and hence (\ref{eq:aQHCR}) holds as $n\to\infty$.
\hfill$\Box$

\subsection{Proofs of proposition~\ref{prop:qcr-multi} and Theorem~\ref{theorem:qhcrk-multi}}
Let 
\begin{align*}
 \langle A,\,B\rangle_{1,\rho}
      &=\frac{1}{2}\tr\rho(BA^\dagger +A^\dagger B),\quad
 \langle A,\,B\rangle_{2,\rho}
      =\frac{1}{2}\tr\rho(BA^\dagger),\\ 
 \vec{X}&=(X^1,X^2,...,X^m)^T,\quad
 Z_\rho(\vec{X})=[\langle X^i, X^j\rangle_{2,\rho}].
\end{align*}
Observe that,
\begin{align*}
  \Re Z_\rho(\vec{X})=[\langle X^i, X^j\rangle_{1,\rho}]. 
\end{align*}

\begin{lemma}
\label{lemma:vz}
\begin{align*}
 V_{\theta}(M)\geq Z_\rho(\vec{X}),
 \quad\quad
 V_{\theta}(M)\geq \Re Z_\rho(\vec{X}).
\end{align*} 
\end{lemma}
\begin{lemma}
Let $\langle \cdot,\cdot\rangle$ be an inner product,
and  $\vec{X}=(X^1,X^2,...,X^m)^T$, 
$\vec{L}=(L_1,L_2,...,L_m)^T$, where 
$X^1,X^2,...,X^m,L_1,L_2,...,L_m $ are 
members of the vector space of our concern.
Denote matrices 
$[\langle X^i, X^j\rangle]$,
and $[\langle L_i, L_j\rangle]$,
by $\Gamma_{\vec{X}}$ and $\Gamma_{\vec{L}}$, 
respectively,
and assume $[\langle X^i, L_j\rangle]=\delta^i_j$.
Then, $\Gamma_{\vec{X}}^{-1}$ and $\Gamma_{\vec{L}}^{-1}$
exists, and 
\begin{align*}
\Gamma_{\vec{X}}\geq 
 \Gamma_{\vec{L}}^{-1}
\end{align*}
holds.
\label{lem:gram}
\end{lemma}
\begin{lemma}
 For any positive hermitian matrix $Z$, we have
\begin{align}
\min_V {\rm Sp} GV={\rm Sp} GZ+{\rm Spabs}\Im GZ,
\end{align} 
where $V$ runs over all positive real matrices such that $V\geq Z$.
\label{lemma:trgv}
\end{lemma}

For the proofs of Lemmas~\ref{lemma:vz}-\ref{lemma:trgv},
see \cite{holevo}.

Let $\langle\cdot,\cdot\rangle$ be $\langle\cdot,\cdot\rangle_{1,\rho_\theta}$
and let $L_i$ be $L^S_{i,\theta}$ (resp. $L^{S,{\bf t}}_{i,\theta,\mbi{\delta}}$ )
in Lemma~\ref{lem:gram}, and combine with Lemma~\ref{lemma:vz} and
Lemma~\ref{lemma:trgv}.
Noticing that $\Im J^S_\theta=0$
(resp. $\Im J^{S,{\bf t}}_{\theta,\mbi{\delta}}=0$), then, we have the first inequality in 
Proposition~\ref{prop:qcr-multi}
(resp. the first inequality in Theorem~\ref{theorem:qhcrk-multi}). 
Similarly,
letting  $\langle\cdot,\cdot\rangle$ be $\langle\cdot,\cdot\rangle_{2,\rho_\theta}$
and letting $L_i$ be $L^R_{i,\theta}$ 
(resp. $L^{R,{\bf t}}_{i,\theta,\mbi{\delta}}$)
in Lemma~\ref{lem:gram}, we have the second inequality in 
Proposition~\ref{prop:qcr-multi} 
(resp. the second inequality in Theorem~\ref{theorem:qhcrk-multi}).

\subsection{Derivation of the equation~(\ref{eq:trace})}
Due to $ \rho^G_{0}=(1-c)\sum_{n=0}^\infty c^{-n},$
where $\frac{c^{-1}}{(1-c^{-1})^2}=\sigma^2-\frac{1}{2},$
we have
\begin{align*}
& \tr\rho^{G-1}_0\rho^G_{(x,y)}\rho^G_{(z,w)}\\
&= (1-c)
\int_{\alpha\in\mathbb{C},\beta\in\mathbb{C}}
\frac{{\rm d}^2\alpha {\rm d}^2\beta}{4\pi^2(\sigma-1/2)^2}
e^{-\frac{|\alpha-x-\sqrt{-1}y|^2+|\beta-z-\sqrt{-1}w|^2}{2(\sigma^2-1/2)}}
\sum_{n=0}^\infty c^n 
\bra{n}\alpha\rangle\bra{\alpha}\beta\rangle\bra{\beta}n\rangle\\
&=(1-c)
\int_{\alpha\in\mathbb{C},\beta\in\mathbb{C}}
\frac{{\rm d}^2\alpha {\rm d}^2\beta}{4\pi^2(\sigma-1/2)^2}
e^{-\frac{|\alpha-x-\sqrt{-1}y|^2+|\beta-z-\sqrt{-1}w|^2}{2(\sigma^2-1/2)}}
\sum_{n=0}^\infty
 c^n\frac{\alpha^n}{\sqrt{n!}}e^{-\frac{1}{2}|\alpha|^2}
e^{-\frac{|\alpha|^2}{2}+\alpha^*\beta-\frac{|\beta|^2}{2}} 
 \frac{\beta^{*n}}{\sqrt{n!}}e^{-\frac{1}{2}|\beta|^2}\\
&=(1-c)
\int_{\alpha\in\mathbb{C},\beta\in\mathbb{C}}
\frac{{\rm d}^2\alpha {\rm d}^2\beta}{4\pi^2(\sigma-1/2)^2}
e^{-\frac{|\alpha-x-\sqrt{-1}y|^2+|\beta-z-\sqrt{-1}w|^2}{2(\sigma^2-1/2)}
  -|\alpha|^2-|\beta|^2+\alpha^*\beta}
\sum_{n=0}^\infty
 \frac{(c\alpha\beta^*)^n}{n!}\\
&=(1-c)
\int_{\alpha\in\mathbb{C},\beta\in\mathbb{C}}
\frac{{\rm d}^2\alpha {\rm d}^2\beta}{4\pi^2(\sigma-1/2)^2}
e^{-\frac{|\alpha-x-\sqrt{-1}y|^2+|\beta-z-\sqrt{-1}w|^2}{2(\sigma^2-1/2)}
  -|\alpha|^2-|\beta|^2+\alpha^*\beta+c\alpha\beta^*}.
\end{align*}
Repeated application of the formula, 
$\int e^{-A(x-B)^2}=\sqrt{\frac{\pi}{A}}$ leads to 
the equation~(\ref{eq:trace}).

\section{Discussions}
\label{sec:discuss}

We have given the first treatment of quanta singular statistical models.
Namely,
we proposed two kinds of quantum versions of HCRK inequality
and Koike inequality.
One is based on difference version of SLD, and the other is
based on the difference version of RLD.
The asymptotic theory based on RLD QHCRK inequality is
also discussed.

We applied those results to some examples.
First example was estimation of entanglement measure,
which is very simple but practically of some importance.
We constructed asymptotically optimal estimate in this case.
Second example is estimation of discrete parameter.
It should be stressed that from the viewpoint of conventional 
state detection theory, the non-asymptotic analysis 
is very hard, while our approach is technically more tractable.
Third example is estimation of vector valued parameter,
and here we could observe the effect of non-commutativity.

There are many unsolved problems.
Especially, asymptotic theory is far from complete.
For example, an asymptotic theory based on SLD QHCRK inequality
is desired, for SLD-based lower bounds are better than
those based on RLD, if the parameter is scalar.
As for the piece-wise differentiable models,
we can trivially extend the theory of differentiable
by imposing asymptotic unbiasedness instead of $\sqrt{n}$-unbiasedness
on the estimator. 
However, SLD versions of
Theorems \ref{th:AQHCRD} and \ref{th:AQHCRC}
are very hard to obtain, for we don't have the equivalence of
(\ref{eq:3:tohi}). Note that this difficulty is due to
non-commutativity, or more specifically,
the hardness of calculation of SLD.
Hence, the similar problems never arise in classical estimation theory.

\subsection*{Acknowledgement}
These results have been obtained when the
first author was a researcher of Imai Quantum Computation
and Information Project, ERATO, JST.
The authors thank
Professor Hiroshi Imai
and
Dr Masahito Hayashi
for comments, helpful discussions, and supports.

\end{document}